\documentclass[preprintnumbers,amsmath,amssymb,floatfix,prd]{revtex4}

\usepackage{graphicx}
\usepackage{dcolumn}
\usepackage{bm}

\newcommand{\beq}{\begin{equation}}
\newcommand{\eeq}{\end{equation}}
\newcommand{\bey}{\begin{eqnarray}}
\newcommand{\eey}{\end{eqnarray}}

\begin{document}                             

\title{Holographic  Cosmology from a System of M2-M5  Branes}

\author{Alireza Sepehri $^{1,2}$\footnote{alireza.sepehri@uk.ac.ir} ,
Mir Faizal $^{3}$ Mohammad Reza Setare
$^{4}$\footnote{rezakord@ipm.ir}, \footnote{f2mir@uwaterloo.ca},
Ahmed Farag Ali$^{5,6}$\footnote{ afali@fsu.edu ;
ahmed.ali@fsu.edu.eg } }
\address{$^1$ Faculty of Physics, Shahid Bahonar University, P.O. Box 76175, Kerman, Iran.\\$^{2}$
Research Institute for Astronomy and Astrophysics of Maragha
(RIAAM), Maragha, Iran.\\$^{3}$ Department of Physics and
Astronomy, University of Waterloo, Waterloo, ON, N2L 3G1,
Canada.\\$^4$  Department of Science, Campus of Bijar, University
of Kurdistan, Bijar, Iran.
\\ $^5$  Department of Physics, Florida State University,
Tallahassee, FL 32306,  USA.
\\ $^6$  Deptartment of Physics, Faculty of Science, Benha University, Benha 13518, Egypt.
}
\begin{abstract}
In this paper, we analyze the holographic cosmology using a M2-M5 brane configuration. In this configuration, a M2-brane will be placed in between a M5-brane and an anti-M5-brane. The M2-brane will act as a channel for energy to flow from an anti-M5-brane to a M5-brane, and this will  increase the degrees of freedom on the M5-brane causing inflation. The inflation will end when the M5-brane and anti-M5-brane get separated. However, at a later stage the distance between the M5-brane and the anti-M5-bran can reduce and this will cause the formation of tachyonic states. These tachyonic states will be again open a bridge between the M5-branes and the anti-M5-branes, which will cause further acceleration of the universe.
\end{abstract}

\maketitle
\section{Introduction}

It is widely known that the entropy of the black holes is proportional to the area of the horizon, and its temperature is proportional to the surface gravity.
Thus, a link between gravity and thermodynamics has been established. This link has become the basis of the  Jacobson formalism  \cite{4}.
In the Jacobson's formalism, the Einsteins equations are   obtained from the first law of thermodynamics. It has been possible to derive the
Friedmann equations from the Clausius relation using this Jacobson formalism  \cite{zza1}. In deriving the  Friedmann equations,  the entropy
is assumed to be proportional to the area of the cosmological horizon. In fact, motivated by the  Jacobson formalism, it has been proposed that
the gravity is an entropic force \cite{zza4}. In fact, Einstein's equations have been obtained from  this entropic force formalism. Furthermore,
as the entropy of the bulk of the black hole is proportional to the area of the boundary, it has led to the development of the holographic principle.
The holographic principle states that the number of degrees of freedom of a region in space is the same as the number of degrees of freedom on the boundary
of that region.

The holographic principle have motivated  the development of the
holographic cosmology \cite{1}-\cite{2012gx}.
The holographic cosmology is based on the idea that the difference between the degrees of freedom
in a region  and the degrees of freedom on the boundary surrounding that region   drives the expansion
of the universe. The holographic cosmology  has been studied in the context of  Lovelock gravity, with
special emphasis on the  Gauss-Bonnet   gravity \cite{9}. The holographic cosmology for the brane world
models \cite{b0}, scalar-tensor gravity \cite{b1}, and $f(R)$ \cite{b2},  has also been discussed.
It may be noted that this analysis was performed using a thermodynamic description of the brane world
models, holographic scalar-tensor cosmology, and holographic $F(R)$ cosmology \cite{f1}. The holographic
cosmology has also been generalized to   the Friedmann-Robertson-Waker universe with an arbitrary spatial curvature
\cite{f2}. This generalization has been  performed for non-flat universes by using the aerial volume instead
of the proper volume \cite{f6}. In this context, the Friedmann equation  for  the Lovelock gravity  with an
arbitrary spatial curvature have also been studied  \cite{f4}.

The holographic cosmology has  been studied using  the BIonic solution \cite{f8}.
This BIon solution is a configuration of a D3-brane and an anti-D3-branes with a wormhole
in between them.   This action for the D-branes is a non-linear action called the
 Dirac-Born-Infeld (DBI) action, and the   non-linearity of this action is important in constructing
 this BIonic solution \cite{b}-\cite{c}.
The F-string end on a point  of a D-brane in this BIonic solution, and the
  F-string charge gets associated with the
world-volume electric flux carried by the D-brane.
 The D3-branes of the BIon has been identified with our universe  and the BIon solution has been used
 for analyzing the holographic cosmology \cite{m}-\cite{mf}.
In this solution. first a BIon forms from black F-strings. Then the degrees of freedom of the D3-brane
increase as energy flows from a anti-D3-brane into the D3-brane. However, as
the  D3-brane moves away from the anti-D3-brane, then the spike of the  D3-brane
 gets  separated from the
 spike of the   anti-D3-brane spike, and the inflation ends at this stage. Finally, when the D3-brane comes
 close to the anti-D3-brane, a new wormhole forms due to the tachyonic states. This wormhole also
 increases the degrees of freedom on the D3-brane, and this causes late time acceleration of the universe.

  It may be noted there the F1-D3 intersection is U-dual to a system of  M2-M5  branes  \cite{f9}-\cite{f10}.
The  M2-branes intersecting with M5-branes have been analyzed
 in the supergravity regime \cite{d1}-\cite{d2}. This analysis has been done using the blackfold approach. Thus,
 it has been possible to recover the $1/4$-BPS self-dual string solution \cite{d4}-\cite{d5},
  as a three-funnel solution of an effective five-brane world volume theory \cite{1d}-\cite{2d}.
  The finite temperature effects for
 non-extremal self-dual string solution solutions and wormhole solutions
 interpolating between stacks of M5-branes and anti-M5-branes have also been studied.
 These solutions define a BIon solution in M-theory  \cite{w}-\cite{w1}.
 It would be interesting to perform a similar analysis for this M-theory system.
 So, it is possible to study the holographic cosmology using a M2-M5 brane system. Thus, we
 will first study a M5-brane connected to an anti-M5-brane by a M2-brane. This will cause the
 degrees of freedom to flow into the M5-brane causing inflation. The inflation will end when the M5-brane
 and the anti-M5-brane get separated. However, at a later stage when the M5-brane approaches the anti-M5-brane,
 tachyonic states will be formed. These tachyonic states will form another bridge  between the M5-brane and
 the anti-M5-brane. We will investigate the inflation in this model of M2-M5 branes, along with the consequences
 of the formation of these tachyonic states. 
 
 It is possible for the four dimensional universe to emerge from compactification of M5-branes. However, in this paper, we will not discuss
such a compactification, and we will  discuss the inflation using M5-brane geometry.
In fact, inflation has already been studied using the M5-brane  geometry \cite{infla}-\cite{infla1}.
 However, in this paper, we study the inflation in the M-theory using the recently proposed proposal of
 holographic cosmology  \cite{1}-\cite{2012gx}.
It may be noted that it is possible for the branes moving in the extra dimensional bulk to collide with each other.
Such collisions have been studied  in the context of  Ekpyrotic universe \cite{coill}-\cite{colli}.
In this paper, we will also use the formalism of holographic cosmology  \cite{infla}-\cite{infla1},
to discuss the state of the universe before such a collision occurs. It may be noted that such a model has also
been studied using  BIonic solution \cite{m}-\cite{mf}, and such a system is U-dual  to a system of  M2-M5  branes  \cite{f9}-\cite{f10}.
So, this system  is actually U-dual to the  BIonic cosmology. In other words, this is the M-theory version of the
holographic cosmology, which has so far only been studied in the string theory   \cite{m}-\cite{mf}.

 We would like to point out that the fact that F1-D3 intersection is U-dual to a system of  M2-M5  branes is only mentioned as an observation at this stage. 
The analysis in this paper seems to be similar to the analysis performed for using BIonic solution, so there might be a deeper 
relation between such a duality. However,  at this stage we only mention this as an motivation to study inflation using 
a M2-M5-brane system. 
We would also like to point out that the bulk is populated by M5-branes and anti-M5-branes. So, a random 
collision between a M5-brane and an anti-M5-brane can occur. However, this will be a collision between a random 
M5-brane with an random anti-M5-brane in the bulk.  Such collision between random branes occurs in  Ekpyrotic universe  \cite{coill}-\cite{colli}. 
However, in this model, we can calculate the state of the universe just before such an collision occurs. 

The paper is organized as the follows.  In section
\ref{o1}, we discuss the   holographic inflationary cosmology using the system of M2-M5 brane.
In
section \ref{o2}, we analyze  the tachyonic states the M2-M5 brane system.
Finally, in the last section, we will summarize our main results.
We will also discuss possible extension of the results obtained in this paper.

\section{ M2-M5 BIonic Solutions }\label{o1}

In this section, we analyze the holographic cosmology using M2-M5 BIonic solutions. This will be done by first analyzing the formation of a configuration of a M5-branes and an anti-M5-brane separated by a M2-brane. Now we can write the
supergravity solution for  black M2-brane lying along $z$ and $r$ directions as follows, \cite{g1}-\cite{1g},

\begin{eqnarray}
 ds^{2} &=& H_{2}^{1/3}[2H_{2}^{-1}du(dv+fdu)+H_{2}^{-1}dz^{2}+\Sigma_{i=1}^{8} dx_{i}^{2}],\nonumber\\
  u &=&-(t-r)/\sqrt{2},\: v = (t+r)/\sqrt{2},\nonumber\\
  H_{2} &=& 1 +
\frac{r_{0}^{3}\sinh^{2}\alpha}{r^{3}},\nonumber \\ f&=&1-\frac{r_{0}^{3}}{r^{3}}.
\label{m1}
\end{eqnarray}
Using this  the definitions, we can write the following expression
\begin{eqnarray}
&&\cosh\alpha_{\pm} = \frac{k\beta^{3}}{\sqrt{2}q_{2}}\frac{\sqrt{1
\pm \sqrt{1 - \frac{4q_{2}^{2}}{k^{2}\beta^{6}}(1 +
\frac{k^{2}}{\sigma^{6}})}}}{\sqrt{1 +
\frac{k^{2}}{\sigma^{6}}}}\nonumber \\&& r_{0,\pm} =
\frac{\sqrt{2}q_{2}}{k\beta^{2}}\frac{\sqrt{1 +
\frac{k^{2}}{\sigma^{6}}}}{\sqrt{1 \pm \sqrt{1 -
\frac{4q_{2}^{2}}{k^{2}\beta^{6}}(1 + \frac{k^{2}}{\sigma^{6}})}}}
\nonumber \\&& \beta = \frac{3}{4\pi T} ,\, q_{2} =
\sigma^{3}\frac{r_{0}^{3}}{2}\sin\theta \sinh2\alpha ,\, \tan\theta = \frac{k}{\sigma^{3}}.
 \label{m2}
\end{eqnarray}
In above equation, $T$ is temperature of M2-brane and $\sigma$
is the radius on the world-volume. The
  mass density along the $z$
direction  is given by
\begin{eqnarray}
&&\frac{M_{M2-brane}}{L_{x^{1}}L_{z}}= Q_{2}\left(1+
\frac{\sqrt{q_{2}}}{3\sqrt{2}\beta^{3}}+\frac{5q_{2}}{2^{6}\beta^{6}}\right)
\label{m3}
\end{eqnarray}
where $Q_{2}=\frac{N_{2}}{(2\pi)^{2}L_{P}^{3}}$. Furthermore, the
  length along  the $x$ and $z$ directions is denoted by
$L_{x}$ and $L_{z}$, respectively. To describe the configuration of a M5-brane and
an anti-M5-brane joined together by an M2-brane, we use  embedding of  the M5-brane
 in $11D$ Minkowski spacetime with metric
\cite{w}-\cite{w1},
\begin{eqnarray}
&& ds^{2} = -dt^{2} +  (dx^{1})^{2} + dr^{2} +
r^{2}d\Omega_{3}^{2} + \sum_{i=6}^{10}dx_{i}^{2}. \label{m4}
\end{eqnarray}
Here we have not considered the  background fluxes. Using the standard angular coordinates
$(\psi,\phi,\omega)$, it is possible to write the following expression,
\begin{eqnarray}
&& d\Omega_{3}^{2} = -d\psi^{2} +  \sin^{2}\psi ( d\phi^{2} +
\sin^{2}\phi d\omega^{2}). \label{m5}
\end{eqnarray}
In  the static gauge, we obtain
\begin{eqnarray}
&&t(\sigma^{a}) = \sigma^{1},\,x^{1}(\sigma^{1}) =
\sigma^{1},\,r(\sigma^{a})=\sigma^{2}\equiv \sigma  \nonumber \\&&
\psi(\sigma^{a}) = \sigma^{3} ,\, \phi(\sigma^{a}) = \sigma^{4},\,
\omega(\sigma^{a}) = \sigma^{5},\, x^{6}(\sigma^{a}) =
z(\sigma), \label{m6}
\end{eqnarray}
and the remaining coordinates are constant. The
embedding function $z(\sigma)$  describes the bending of the
brane. As $z$ is  a transverse coordinate to the branes and $\sigma$
is the radius on the world volume, we can write the
induced metric on the effective five-brane world volume as
\begin{eqnarray}
\gamma_{ab}d\sigma^{a}d\sigma^{b} = -(d\sigma^{0})^{2} +
(d\sigma^{1})^{2} + (1 + z'(\sigma)^{2})d\sigma^{2} +
\sigma^{2}(-d\psi^{2} +  \sin^{2}\psi ( d\phi^{2} + \sin^{2}\phi
d\omega^{2})) \label{m7}
\end{eqnarray}
Here we have imposed two boundary
 conditions in our system. Thus, we have imposed  $z(\sigma)\rightarrow 0$ for
 $\sigma\rightarrow \infty$ and $z'(\sigma)\rightarrow -\infty$ for $\sigma\rightarrow \sigma_{0}$,
 where $\sigma_{0}$ is the minimal two-sphere radius of the
configuration. The mass density along the $z$
direction  at corresponding point can now be expressed as  \cite{w}-\cite{w1},
\begin{eqnarray}
&& \frac{dM_{BIon}}{L_{x^{1}}dz}=
Q_{2} \left(\sqrt{1+\frac{\sigma^{6}}{k^{2}}} +
\frac{5q_{2}^{2}}{6\beta^{6}}\frac{\left(1+\frac{\sigma^{6}}{k^{2}}\right)^{3/2}}{\sigma_{0}^{6}}
+
\frac{11q_{2}^{4}}{8\beta^{12}}\frac{\left(1+\frac{\sigma^{6}}{k^{2}}\right)^{5/2}}{\sigma_{0}^{12}}
\right)\label{m8}
\end{eqnarray}
where we have used of this fact that  $q_{2}=kq_{5}$.
We now compare the mass densities for M2-M5 system to the mass density
for black M2-brane and note that  the  thermal BIon at
$\sigma = \sigma_{0}$  behaves like black M2-brane. Furthermore, minimal two-sphere radius of the
configuration  $\sigma_{0}$
depends  on the temperature
\cite{w}-\cite{w1},
\begin{eqnarray}
&& \sigma_{0} = \frac{q_{2}^{1/4}}{\beta^{1/2}}\left(1.234 -
0.068\frac{q_{2}^{1/2}}{\beta^{3}}\right) \label{m9}
\end{eqnarray}
 The inflation ends when the the M5-brane gets separated from the
 anti-M5-brane, and this happens when the   width of geometry between them reaches zero,
\begin{eqnarray}
&& \sigma_{0}=0,
\beta = \frac{3}{4\pi T}\rightarrow T_{end}=\frac{3}{4\pi}(\frac{0.05}{q_{2}^{1/2}})^{2/7}.
\label{m10}
\end{eqnarray}
Thus, the   temperature was infinity at the
beginning, and then it  decreased   to $T_{end}$
at the end of inflation. This corresponded to the decrease of the
width of the geometry between M5-brane and
the anti-M5-brane till it reached  zero. This corresponded to the limiting value of the
temperature was $T=T_{end}$.

This M2-brane geometry  acted as a channel for the degrees of freedom to flow into   the
 M5-brane, and this in turn led to
  inflation. It may be noted that by
putting the black M2-brane
charge along the radial direction and using equation (\ref{m7}),
we obtain \cite{w}-\cite{w1},
\begin{eqnarray}
z_{\pm}(\sigma)= \int_{\sigma}^{\infty} ds
\left(\frac{F_{\pm}(s)^{2}}{F_{\pm}(\sigma_{0})^{2}}-1\right)^{-\frac{1}{2}}
\label{m11}
\end{eqnarray}
It is possible to express $F(\sigma)$ for the  finite temperature M2-M5 brane system as
\begin{eqnarray}
F_{\pm}(\sigma) = \sigma^{3}\left(\frac{1 + \frac{k^{2}}{\sigma^{6}}}{1
\pm \sqrt{1 - \frac{4q_{5}^{2}}{\beta^{6}}(1 +
\frac{k^{2}}{\sigma^{6}})}}\right)^{3/2}\left(-2 +
\frac{3\beta^{6}}{2q_{5}^{2}}\frac{1 \pm \sqrt{1 -
\frac{4q_{5}^{2}}{\beta^{6}}(1 + \frac{k^{2}}{\sigma^{6}})}}{1 +
\frac{k^{2}}{\sigma^{6}}}\right) \label{m12}
\end{eqnarray}
where
\begin{eqnarray}
&&q_{5} =
\frac{r_{0}^{3}}{2}\cos\theta \sinh2\alpha
\nonumber \\&& \tan\theta = \frac{k}{\sigma^{3}},\,q_{2} = k q_{5}
= -4\pi \frac{N_{2}}{N_{5}}l_{p}^{3}
 \label{m13}
\end{eqnarray}
Here the number of M2-branes  and
M5-branes  is denoted by  $N_{2}$ and $N_{5}$, and the charges on these
branes is denoted by $q_{2}$ and $q_{5}$,  respectively. The temperature of this system is
denoted by $T$.   Attaching a
mirror solution to Eq. (\ref{m11}), we construct thin shell  wormhole
configuration. Now we find an expression for the distance separating the
$N$ M5-branes and   $N$ anti-M5-branes. This is done by defining
  $\Delta = 2z(\sigma_{0})$, where
\begin{eqnarray}
\Delta = 2z(\sigma_{0})= 2\int_{\sigma_{0}}^{\infty}
ds\left(\frac{F(s)^{2}}{F(\sigma_{0})^{2}}-1\right)^{-\frac{1}{2}}.
\label{m14}
\end{eqnarray}
We can now use these results for constructing the holographic cosmology in the    M2-M5 BIon.
In order to do that  we
need to compute the contribution of the M2-M5 system to the
 the surface degrees of freedom on the holographic
horizon and the bulk degrees of freedom inside the universe. We mean by the bulk  the region inside the cosmological horizon of a universe,
and it should not be confused with the space in which  the M2-M5 branes geometry
Now  the  relations between these degrees
of freedom,  the entropy of M2-M5,  and  the mass density
along the transverse direction, can be written as
 \begin{eqnarray}
&&N_{sur} + N_{bulk} = N_{M2-M5} = N(M5-brane) + N(anti-M5-brane)
+ N(M2-brane) \nonumber \\&&\simeq
4L_{P}^{2}S_{M2-M5}=\frac{\Omega_{(3)}\Omega_{(4)}}{16\pi G}\int
d\sigma\frac{F(\sigma)}{F(\sigma_{0})}\beta^{4}\sigma^{3}\frac{1}{\cosh^{3}\alpha}
\nonumber \\&& N_{sur} - N_{bulk} \simeq \int d\sigma
\frac{dM_{M2-M5}}{dz}=\frac{\Omega_{(3)}\Omega_{(4)}}{16\pi G}\int
d\sigma
\frac{F(\sigma)}{F(\sigma_{0})}\beta^{3}\sigma^{3}\frac{3\cosh^{2}\alpha
+ 1}{\cosh^{3}\alpha} \label{m15}
\end{eqnarray}
where $\Omega_{n}$ denotes the volume of the round $N$-sphere.
The solution to these equation, can be written as
\begin{eqnarray}
&&N_{sur} \simeq \frac{\Omega_{(3)}\Omega_{(4)}}{16\pi G}\int
d\sigma\frac{F(\sigma)}{F(\sigma_{0})}\beta^{4}\sigma^{3}\frac{1}{\cosh^{3}\alpha}
+ \frac{\Omega_{(3)}\Omega_{(4)}}{16\pi G}\int d\sigma
\frac{F(\sigma)}{F(\sigma_{0})}\beta^{3}\sigma^{3}\frac{3\cosh^{2}\alpha
+ 1}{\cosh^{3}\alpha} \nonumber \\&&  N_{bulk} \simeq
\frac{\Omega_{(3)}\Omega_{(4)}}{16\pi G}\int
d\sigma\frac{F(\sigma)}{F(\sigma_{0})}\beta^{4}\sigma^{3}\frac{1}{\cosh^{3}\alpha}
- \frac{\Omega_{(3)}\Omega_{(4)}}{16\pi G}\int d\sigma
\frac{F(\sigma)}{F(\sigma_{0})}\beta^{3}\sigma^{3}\frac{3\cosh^{2}\alpha
+ 1}{\cosh^{3}\alpha} \label{m16}
\end{eqnarray}
The temperature of the BIonic system decreases  with the increase in the
degrees of freedom inside the universe.

It is also possible to express the
 Hubble parameter and energy density
in terms of the quantities associated with the M2-M5 brane system. As the number of degrees
of freedom on the   apparent horizon
 is proportional to its area, we can write
\begin{eqnarray}
&& N_{sur} = \frac{4\pi r_{A}^{2}}{L_{P}^{2}}+\frac{8}{3}\alpha_{0}\sqrt{\frac{\pi\mu}{L_{P}^{2}}}r_{A}\label{m17},
\end{eqnarray}
where  the  $H$ is the Hubble parameter which can  be expressed in terms of the
 scale
factor $a$ as $H=
\frac{\dot{a}}{a}$. The radius $r_{A} = 1/ \sqrt{H^{2} + \frac{\bar{k}}{a^{2}}}$ is  the
apparent horizon radius for the  Universe.
The Hubble parameter
for flat universe can now be expressed as
\begin{eqnarray}
&& H_{flat,inf} \simeq \Big(\frac{16\pi G k^{4}
q_{2}^{6}}{3\Omega_{(3)}\Omega_{(4)}q_{5}^{6}\sigma_{0}^{3}} T^{3}
+ \frac{8\pi G k^{2}
q_{2}^{4}}{\Omega_{(3)}\Omega_{(4)}q_{5}^{4}\sigma_{0}^{3}}
T^{2}\Big)\nonumber \\ &\times& \Bigg(\frac{8}{3}\alpha_{0}\sqrt{\frac{\pi\mu}{L_{P}^{2}}}\pm \sqrt{\frac{16}{3}\alpha_{0}^{2}\frac{\pi\mu}{L_{P}^{2}}+\frac{16\pi }{L_{P}^{2}}\Big(\frac{16\pi G k^{4}
q_{2}^{6}}{3\Omega_{(3)}\Omega_{(4)}q_{5}^{6}\sigma_{0}^{3}} T^{3}
+ \frac{8\pi G k^{2}
q_{2}^{4}}{\Omega_{(3)}\Omega_{(4)}q_{5}^{4}\sigma_{0}^{3}}
T^{2}\Big)^{-1}}\Bigg).  \label{m18}
\end{eqnarray}
The universe energy density of the universe can be written as
\begin{eqnarray}
&& \rho_{flat,inf} = \frac{3}{8\pi L_{P}^{2}}H_{flat,inf}^{2} \simeq \frac{3}{8\pi L_{P}^{2}}
\Big(\frac{16\pi G k^{4}
q_{2}^{6}}{3\Omega_{(3)}\Omega_{(4)}q_{5}^{6}\sigma_{0}^{3}} T^{3}
+ \frac{8\pi G k^{2}
q_{2}^{4}}{\Omega_{(3)}\Omega_{(4)}q_{5}^{4}\sigma_{0}^{3}}
T^{2}\Big)^{2}\nonumber \\ &\times& \Bigg(\frac{8}{3}\alpha_{0}\sqrt{\frac{\pi\mu}{L_{P}^{2}}}\pm \sqrt{\frac{16}{3}\alpha_{0}^{2}\frac{\pi\mu}{L_{P}^{2}}+\frac{16\pi }{L_{P}^{2}}\Big(\frac{16\pi G k^{4}
q_{2}^{6}}{3\Omega_{(3)}\Omega_{(4)}q_{5}^{6}\sigma_{0}^{3}} T^{3}
+ \frac{8\pi G k^{2}
q_{2}^{4}}{\Omega_{(3)}\Omega_{(4)}q_{5}^{4}\sigma_{0}^{3}}
T^{2}\Big)^{-1}}\Bigg)^{2} \label{m19}
\end{eqnarray}
The energy density of the universe is related to the
 charges of M-branes. It is also related to the
geometry of the
 M2-M5 system.

It is also possible to use Eq. (\ref{m17}), and
the expression $r_{A} = 1/ \sqrt{H^{2} + \frac{\bar{k}}{a^{2}}}$, to obtain the  Hubble parameter
for non-flat universe as
\begin{eqnarray}
&& H_{o/c,inf} \simeq\Bigg[\Big(\frac{16\pi G k^{4}
q_{2}^{6}}{3\Omega_{(3)}\Omega_{(4)}q_{5}^{6}\sigma_{0}^{3}} T^{3}
+ \frac{8\pi G k^{2}
q_{2}^{4}}{\Omega_{(3)}\Omega_{(4)}q_{5}^{4}\sigma_{0}^{3}}
T^{2}\Big)^{2}\nonumber \\ &\times& \Bigg(\frac{8}{3}\alpha_{0}\sqrt{\frac{\pi\mu}{L_{P}^{2}}}\pm \sqrt{\frac{16}{3}\alpha_{0}^{2}\frac{\pi\mu}{L_{P}^{2}}+\frac{16\pi }{L_{P}^{2}}\Big(\frac{16\pi G k^{4}
q_{2}^{6}}{3\Omega_{(3)}\Omega_{(4)}q_{5}^{6}\sigma_{0}^{3}} T^{3}
+ \frac{8\pi G k^{2}
q_{2}^{4}}{\Omega_{(3)}\Omega_{(4)}q_{5}^{4}\sigma_{0}^{3}}
T^{2}\Big)^{-1}}\Bigg)^{2} -\bar{K}/a^{2}\Bigg]^{1/2}\label{m20}.
\end{eqnarray}
The scale factor for open $(k=-1)$ universe can be written as
\begin{eqnarray}
&& a_{o,inf}(t) \simeq \exp -\int dt [ \mathcal{T}  + ln(t)  ]\label{m22}.
\end{eqnarray}
The  scale factor for closed $(k=+1)$ universe can be written as
\begin{eqnarray}
&& a_{c,inf}(t) \simeq \exp -i\int dt [  \mathcal{T} + ln(t) + \frac{\pi}{2}  ]\label{m23}.
\end{eqnarray}
Here we have defined
  the quantity $ \mathcal{T}$ as
\begin{eqnarray}
 \mathcal{T} &=& \Big(\frac{16\pi G k^{4}
q_{2}^{6}}{3\Omega_{(3)}\Omega_{(4)}q_{5}^{6}\sigma_{0}^{3}} T^{3}
+ \frac{8\pi G k^{2}
q_{2}^{4}}{\Omega_{(3)}\Omega_{(4)}q_{5}^{4}\sigma_{0}^{3}}
T^{2}\Big)^{2}\nonumber \\ &\times& \Bigg(\frac{8}{3}\alpha_{0}\sqrt{\frac{\pi\mu}{L_{P}^{2}}}\pm \sqrt{\frac{16}{3}\alpha_{0}^{2}\frac{\pi\mu}{L_{P}^{2}}+\frac{16\pi }{L_{P}^{2}}\Big(\frac{16\pi G k^{4}
q_{2}^{6}}{3\Omega_{(3)}\Omega_{(4)}q_{5}^{6}\sigma_{0}^{3}} T^{3}
+ \frac{8\pi G k^{2}
q_{2}^{4}}{\Omega_{(3)}\Omega_{(4)}q_{5}^{4}\sigma_{0}^{3}}
T^{2}\Big)^{-1}}\Bigg)^{2}. \label{m21}
\end{eqnarray}
The scale factor for the open universe is very small at the  the beginning ($T=\infty$).
However,   as the temperature decreases
the value for the scalar factor increases. Thus, at the end of the inflation, this scalar factor has a large
value.  It may be noted that unlike a open universe,   the scale factor for a closed
 universe oscillates. The
 energy density for open and closed universes,  can now be expressed as follows,
 \begin{eqnarray}
  \rho_{o/c,inf} &=& \frac{3}{8\pi l_{P}^{2}}\Big(H_{o/c}^{2}+k/a^{2}\Big)  \nonumber \\&\simeq&
   \frac{3}{8\pi l_{P}^{2}}H_{flat}^{2} \nonumber \\ &\simeq&\frac{3}{8\pi L_{P}^{2}}
\Big(\frac{16\pi G k^{4}
q_{2}^{6}}{3\Omega_{(3)}\Omega_{(4)}q_{5}^{6}\sigma_{0}^{3}} T^{3}
+ \frac{8\pi G k^{2}
q_{2}^{4}}{\Omega_{(3)}\Omega_{(4)}q_{5}^{4}\sigma_{0}^{3}}
T^{2}\Big)^{2}\nonumber \\ &\times& \Bigg(\frac{8}{3}\alpha_{0}\sqrt{\frac{\pi\mu}{L_{P}^{2}}}\pm \sqrt{\frac{16}{3}\alpha_{0}^{2}\frac{\pi\mu}{L_{P}^{2}}+\frac{16\pi }{L_{P}^{2}}\Big(\frac{16\pi G k^{4}
q_{2}^{6}}{3\Omega_{(3)}\Omega_{(4)}q_{5}^{6}\sigma_{0}^{3}} T^{3}
+ \frac{8\pi G k^{2}
q_{2}^{4}}{\Omega_{(3)}\Omega_{(4)}q_{5}^{4}\sigma_{0}^{3}}
T^{2}\Big)^{-1}}\Bigg)^{2}\nonumber \\ &=&\rho_{flat,inf}.
\label{m24}
\end{eqnarray}
The energy density of the universes is related to the evolution of the M2-M5 brane system
so it does not depend on type of universe. This can be noted from the fact that the energy density  of the flat, open and closed universes is the same.

The inflation ends when the M5-brane is separated from the anti-M5-brane. At this stage
the mass distribution along $z$-direction is absent. Now we can write the difference between the surface and
bulk degrees of freedom as
\begin{eqnarray}
&& N_{sur} - N_{bulk} \simeq \int_{\sigma_{0}}^{\sigma_{0}}
d\sigma \frac{dM_{BIon}}{dz}=0 \label{m25},
\end{eqnarray}
and so we have
\begin{eqnarray}
 N_{sur} &=& N_{bulk} \label{m26}.
\end{eqnarray}
The   degrees of freedom   on the cosmological horizon is equal to the degrees of freedom in the
  bulk, at the end of inflation.

\section{Tachyonic States }\label{o2}
 In the previous section we analysed the
 inflation in the context of M2-M5 brane system. The inflation ended with the M5-brane
 getting separated from the anti-M5-brane. However, it is possible for the
 M5-brane to again come close to the anti-M5-brane at a later stage.
 This can result in a collision of the M5-brane with the anti-M5-brane.
It may be noted that such a collision of branes has been studied in the context of Ekpyrotic universe \cite{coill}-\cite{colli}.
In this section, we will analyse the state of the universe before such a collision using the formalism of holographic cosmology \cite{1}-\cite{2012gx}.
So,
 we will analyse phenomena of a M5-brane approaching an anti-M5-brane. It will be demonstrated that
 this will create tachyonic states, and these tachyonic states will in turn create a new bridge between the
 M5-brane and the anti-M5-brane. This may be responsible for the expansion
 at a later stage in the evolution of the universe.
 It may be noted here the universe starts with a
  non-phantom phase and then evolves to a
phantom one.

The non-phantom phase can be constructed by  using a
M5-branes and an anti-M5-branes in the background  (\ref{m3}). Thus, in this model a M5-brane moving in the extra dimension approaches
an anti-M5-brane, and   tachyonic states form when
the distance between the M5-brane and the anti-M5-brane reaches a critical distance $l$.
So, now we will analyse this critical system by placing one of the branes at
 $z_{1} = l/2$,  and the other brane   at $z_{2} = -l/2$. At this point the
tachyonic states will form and the universe will enter a near collapse phase. We will use the formalism of
 holographic cosmology \cite{1}-\cite{2012gx} for analysing this near collapse phase.
  So, the separation between the M5-brane and the anti-M5-branes becomes of the order one.
  We need a tachyonic action to analyse this state.    Now we can write the
open string tachyon for this system as   \cite{q2}-\cite{m2},
\begin{eqnarray}
&& H_{DBI} =\frac{\Omega_{3}\Omega_{4}L_{t}L_{x^{1}}}{16\pi G}\frac{2^{3/2}q_{5}^{3}}{\beta^{6}} \int d\sigma V(TA)\left(\sqrt{1 + \frac{l'(\sigma)^{2}}{4}
+ \dot{TA}^{2} -  TA'^{2}}\right)F_{DBI,M2-M5} ,  \nonumber \\&&
F_{DBI,M2-M5}=\sigma^{3}\left(\frac{1 + \frac{k^{2}}{\sigma^{6}}}{1
\pm \sqrt{1 - \frac{4q_{5}^{2}}{\beta^{6}}(1 +
\frac{k^{2}}{\sigma^{6}})}}\right)^{3/2}\left(-2 +
\frac{3\beta^{6}}{2q_{5}^{2}}\frac{1 \pm \sqrt{1 -
\frac{4q_{5}^{2}}{\beta^{6}}(1 + \frac{k^{2}}{\sigma^{6}})}}{1 +
\frac{k^{2}}{\sigma^{6}}}\right)\left(1-\frac{64q_{5}}{162\beta^{3}}\right)\label{m27}.
\end{eqnarray}
Here we have used
 \begin{eqnarray}
V(TA)=\frac{\tau_{3}}{ \cosh\sqrt{\pi}TA} \label{m28}.
\end{eqnarray}
So, we can write the
 equation of motion for $l(\sigma)$ and tachyon $TA$, as
\begin{eqnarray}
&&\left(\frac{l''V(TA)F_{DBI,M2-M5}}{4\sqrt{1+
\frac{l'(\sigma)^{2}}{4}+ \dot{TA}^{2} -  TA'^{2}}}\right)+\nonumber \\&&\left(\frac{l'}{4\sqrt{1+
\frac{l'(\sigma)^{2}}{4}+ \dot{TA}^{2} -  TA'^{2}}}\right)(V'(TA)F_{DBI,M2-M5}+V(TA)F'_{DBI,M2-M5})+\nonumber \\&&
\frac{2l'(\dot{TA}\dot{TA}' -  TA'TA'')}{1 + \frac{l'(\sigma)^{2}}{4}
+ \dot{TA}^{2} -  TA'^{2}}V(TA)F_{DBI,M2-M5}=0
\label{m29}.
\end{eqnarray}
We also have
\begin{eqnarray}
&&\left(\frac{1}{\sqrt{D_{TA,M2-M5}}}TA'(\sigma)\right)'=\nonumber \\&&\frac{1}{\sqrt{D_{TA,M2-M5}}}
\left[\frac{(V(TA))'}{(V(TA))}(D_{TA,M2-M5}-(TA'(\sigma))^{2})+\frac{F'_{DBI,M2-M5}}{F_{DBI,M2-M5}}D_{TA,M2-M5}\right],\nonumber \\&&D_{TA,M2-M5}=1 + \frac{l'(\sigma)^{2}}{4}
+ \dot{TA}^{2} -  TA'^{2}
\label{m30}
\end{eqnarray}
The solution to these equations can be written as
\begin{eqnarray}
&&l(\sigma) = 2l_{0}\left(\frac{1}{2} -\int_{\sigma}^{\infty} d\sigma
\left(\frac{F_{DBI,M2-M5}(\sigma)+4\int d\sigma'(\dot{TA}^{2} -
TA'^{2} )}{F_{DBI,M2-M5}(\sigma_{0})}-1\right)^{-\frac{1}{2}}\right),
\label{m31}
\end{eqnarray}
where
\begin{eqnarray}
TA\sim
\left(\frac{\sigma_{0}^{2}}{\sigma^{2}-\sigma_{0}^{2}}\right)^{1/3}\left(\frac{81\beta^{3}}{192q_{5}}-1\right)\label{m32}.
\end{eqnarray}
The non-vanishing value of
  $\sigma_{0}$  represents a  geometry that connects the M5-brane with the anti-M5-brane, and has a
  finite size width.
The  distance separating the
 M5-brane from the  anti-M5-brane  is $l_{0}$, when this geometry is formed.
It may be noted that there are no tachyonic states present before this geometry forms, and as
the temperature decreases the size of this geometry increases.

  The  entropy and mass density along $z$-direction in this  tachyonic theory
can be written as
 \begin{eqnarray}
 S_{tb}&=& \frac{2^{3/2}q_{5}^{3}\Omega_{3}\Omega_{4}}{16\pi G} \int
d\sigma V(TA(\sigma))\frac{F_{DBI,M2-M5}(\sigma)}{F_{DBI,M2-M5}(\sigma_{0})}\beta^{4}\sigma^{3}\nonumber \\ && \times
\frac{1}{ \cosh^{3}\alpha}\frac{\sigma_{0}^{7/3}}{(\sigma^{2}-\sigma_{0}^{2})^{4/3}}\left(\frac{81\beta^{3}}{192q_{5}}-1\right) \label{m33} \\
\frac{dM_{tb}}{dz}&=&\frac{2^{3/2}q_{5}^{3}\Omega_{3}\Omega_{4}}{4 G}
 V(TA(\sigma))\frac{F_{DBI,M2-M5}(\sigma)}{F_{DBI,M2-M5}(\sigma_{0})}\beta^{3}\sigma^{3}\frac{3 \cosh^{2}\alpha +
1}{ \cosh^{3}\alpha}\nonumber \\ && \times\frac{\sigma_{0}^{7/3}}{(\sigma^{2}-\sigma_{0}^{2})^{4/3}}\left(\frac{81\beta^{3}}{192q_{5}}-1\right) \label{m34}.
\end{eqnarray}

These   tachyonic states also effect the    of degrees of freedom on the
universe and cause further acceleration. Now we can again write the expression for the
 degrees
of freedom in terms of the  the entropy of this system. It is also possible to write an expression for the
mass   mass density
along the transverse direction for this system. Thus, we can write the following expression,
 \begin{eqnarray}
 N_{sur} + N_{bulk} &=& N_{BIon}= N_{brane} + N_{anti-brane} +
N_{wormhole}\nonumber \\&  \simeq &
\frac{2^{3/2}q_{5}^{3}\Omega_{3}\Omega_{4}}{16\pi G} \int
d\sigma V(TA(\sigma))\frac{F_{DBI,M2-M5}(\sigma)}{F_{DBI,M2-M5}(\sigma_{0})}\beta^{4}\sigma^{3}\nonumber \\ && \times
\frac{1}{ \cosh^{3}\alpha}\frac{\sigma_{0}^{7/3}}{(\sigma^{2}-\sigma_{0}^{2})^{4/3}}\left(\frac{81\beta^{3}}{192q_{5}}-1\right)
\nonumber \\  N_{sur} - N_{bulk} &\simeq& \int d\sigma
\frac{dM_{BIon}}{dz}\nonumber \\ &=&
\frac{2^{3/2}q_{5}^{3}\Omega_{3}\Omega_{4}}{4 G}
 V(TA(\sigma))\frac{F_{DBI,M2-M5}(\sigma)}{F_{DBI,M2-M5}(\sigma_{0})}\beta^{3}\sigma^{3}\frac{3 \cosh^{2}\alpha +
1}{ \cosh^{3}\alpha}\nonumber \\ && \times\frac{\sigma_{0}^{7/3}}{(\sigma^{2}-\sigma_{0}^{2})^{4/3}}\left(\frac{81\beta^{3}}{192q_{5}}-1\right)\label{m35}.
\end{eqnarray}
The solution of these equations can be written as
\begin{eqnarray}
  N_{sur} &\simeq& \frac{2^{3/2}q_{5}^{3}\Omega_{3}\Omega_{4}}{16\pi G} \int
d\sigma V(TA(\sigma))\frac{F_{DBI,M2-M5}(\sigma)}{F_{DBI,M2-M5}(\sigma_{0})}\beta^{4}\sigma^{3}\nonumber \\ && \times
\frac{1}{ \cosh^{3}\alpha}\frac{\sigma_{0}^{7/3}}{(\sigma^{2}-\sigma_{0}^{2})^{4/3}}\left(\frac{81\beta^{3}}{192q_{5}}-1\right)
\nonumber \\&&
+ \frac{2^{3/2}q_{5}^{3}\Omega_{3}\Omega_{4}}{4 G}
 V(TA(\sigma))\frac{F_{DBI,M2-M5}(\sigma)}{F_{DBI,M2-M5}(\sigma_{0})}\beta^{3}\sigma^{3}\frac{3 \cosh^{2}\alpha +
1}{ \cosh^{3}\alpha}\nonumber \\ && \times\frac{\sigma_{0}^{7/3}}{(\sigma^{2}-\sigma_{0}^{2})^{4/3}}\left(\frac{81\beta^{3}}{192q_{5}}-1\right), \nonumber \\   N_{bulk}
&\simeq& \frac{2^{3/2}q_{5}^{3}\Omega_{3}\Omega_{4}}{16\pi G} \int
d\sigma V(TA(\sigma))\frac{F_{DBI,M2-M5}(\sigma)}{F_{DBI,M2-M5}(\sigma_{0})}\beta^{4}\sigma^{3}\nonumber \\ && \times
\frac{1}{ \cosh^{3}\alpha}\frac{\sigma_{0}^{7/3}}{(\sigma^{2}-\sigma_{0}^{2})^{4/3}}\left(\frac{81\beta^{3}}{192q_{5}}-1\right)
\nonumber \\&&
- \frac{2^{3/2}q_{5}^{3}\Omega_{3}\Omega_{4}}{4 G}
 V(TA(\sigma))\frac{F_{DBI,M2-M5}(\sigma)}{F_{DBI,M2-M5}(\sigma_{0})}\beta^{3}\sigma^{3}\frac{3 \cosh^{2}\alpha +
1}{ \cosh^{3}\alpha}\nonumber \\ && \times\frac{\sigma_{0}^{7/3}}{(\sigma^{2}-\sigma_{0}^{2})^{4/3}}\left(\frac{81\beta^{3}}{192q_{5}}-1\right) \label{m36}.
\end{eqnarray}
The M5-brane approaches the anti-M5-brane, and  this causes the production of
tachyonic states. These states  increase the  number of degrees of freedom of the universe
and this continues till the
the  Big Rip  singularity. The Hubble parameter
for flat universe can be expressed as
\begin{eqnarray}
&& H_{flat,ac} \simeq \left(\frac{1}{V(TA)}\right)^{1/2}\Big(\frac{1016q_{2}^{3}G\pi^{5}}{243\Omega_{(3)}\Omega_{(4)}q_{5}^{3}\sigma_{0}^{7/3}} T^{5} + \frac{64q_{2}G\pi^{3}}{27\Omega_{(3)}\Omega_{(4)}q_{5}\sigma_{0}^{7/3}}
T^{3}\Big)\nonumber \\ &\times& \Bigg(\frac{8}{3}\alpha_{0}\sqrt{\frac{\pi\mu}{L_{P}^{2}}}\pm \sqrt{\frac{16}{3}\alpha_{0}^{2}\frac{\pi\mu}{L_{P}^{2}}+\frac{16\pi }{L_{P}^{2}}\Big(\frac{1016q_{2}^{3}G\pi^{5}}{243\Omega_{(3)}\Omega_{(4)}q_{5}^{3}\sigma_{0}^{7/3}} T^{5} + \frac{64q_{2}G\pi^{3}}{27\Omega_{(3)}\Omega_{(4)}q_{5}\sigma_{0}^{7/3}}
T^{3}\Big)^{-1}}\Bigg).\label{m37}
\end{eqnarray}
As the    tachyonic potential increases, this Hubble parameter reduces to very small values.

 The   energy density can also  be calculated as follows,
\begin{eqnarray}
  \rho_{flat,ac} &=&  \frac{3}{8\pi L_{P}^{2}}H_{flat,ac}^{2}
  \nonumber \\  &=& \frac{3}{8\pi L_{P}^{2}}\left(\frac{1}{V(TA)}\right)\Big(\frac{1016q_{2}^{3}G\pi^{5}}{243\Omega_{(3)}\Omega_{(4)}q_{5}^{3}\sigma_{0}^{7/3}} T^{5} + \frac{64q_{2}G\pi^{3}}{27\Omega_{(3)}\Omega_{(4)}q_{5}\sigma_{0}^{7/3}}
T^{3}\Big)^{2}\nonumber \\ &\times& \Bigg(\frac{8}{3}\alpha_{0}\sqrt{\frac{\pi\mu}{L_{P}^{2}}}\pm \sqrt{\frac{16}{3}\alpha_{0}^{2}\frac{\pi\mu}{L_{P}^{2}}+\frac{16\pi }{L_{P}^{2}}\Big(\frac{1016q_{2}^{3}G\pi^{5}}{243\Omega_{(3)}\Omega_{(4)}q_{5}^{3}\sigma_{0}^{7/3}} T^{5} + \frac{64q_{2}G\pi^{3}}{27\Omega_{(3)}\Omega_{(4)}q_{5}\sigma_{0}^{7/3}}
T^{3}\Big)^{-1}}\Bigg)^{2}
\label{m40}
\end{eqnarray}
This energy density decreases with increasing
tachyon potential. This occurs because   the   acceleration decreases the energy density.
The acceleration occurs due to the tachyonic states forming a bridge between the M5-brane and the anti-M5-brane.

Now using the radius  $r_{A} = 1/ \sqrt{H^{2} + \frac{\bar{k}}{a^{2}}}$,  the Hubble parameter
for non-flat universes can be written as
\begin{eqnarray}
  H_{o/c,ac} &\simeq& \Bigg[ \left(\frac{1}{V(TA)}\right)\Big(\frac{1016q_{2}^{3}G\pi^{5}}{243\Omega_{(3)}\Omega_{(4)}q_{5}^{3}\sigma_{0}^{7/3}} T^{5} + \frac{64q_{2}G\pi^{3}}{27\Omega_{(3)}\Omega_{(4)}q_{5}\sigma_{0}^{7/3}}
T^{3}\Big)^{2}\nonumber \\ &&\times \Bigg(\frac{8}{3}\alpha_{0}\sqrt{\frac{\pi\mu}{L_{P}^{2}}}\nonumber \\ && \pm \sqrt{\Bigg(\frac{16}{3}\alpha_{0}^{2}\frac{\pi\mu}{L_{P}^{2}}+\frac{16\pi }{L_{P}^{2}}\left(\frac{1016q_{2}^{3}G\pi^{5}}{243\Omega_{(3)}\Omega_{(4)}q_{5}^{3}\sigma_{0}^{7/3}} T^{5} + \frac{64q_{2}G\pi^{3}}{27\Omega_{(3)}\Omega_{(4)}q_{5}\sigma_{0}^{7/3}}
T^{3}\right)^{-1}\Bigg)}\Bigg)^{2}-\bar{K}/a^{2}\Bigg]^{1/2}\label{mm19}.
\end{eqnarray}

The scale factor for open   universe can now be expressed as
\begin{eqnarray}
a_{o,ac}(t) \simeq \exp-\int dt  [ \mathcal{T}_t + ln(t) ],&& \label{mmm19}
\end{eqnarray}
The scale factor for  closed universe  can also be expressed as
\begin{eqnarray}
 a_{c,ac}(t) \simeq \exp -i\int dt  [ \mathcal{T}_t + ln(t) + \frac{\pi}{2}  ] .&& \label{mmmm19}
\end{eqnarray}
Here we have defined
$\mathcal{T}_t$ as
\begin{eqnarray}
 \mathcal{T}_t &=& \left(\frac{1}{V(TA)}\right)\Big(\frac{1016q_{2}^{3}G\pi^{5}}{243\Omega_{(3)}\Omega_{(4)}q_{5}^{3}\sigma_{0}^{7/3}} T^{5} + \frac{64q_{2}G\pi^{3}}{27\Omega_{(3)}\Omega_{(4)}q_{5}\sigma_{0}^{7/3}}
T^{3}\Big)^{2}\nonumber \\ &\times& \Bigg(\frac{8}{3}\alpha_{0}\sqrt{\frac{\pi\mu}{L_{P}^{2}}}\pm \sqrt{ \Big(\frac{16}{3}\alpha_{0}^{2}\frac{\pi\mu}{L_{P}^{2}}+\frac{16\pi }{L_{P}^{2}}\Big(\frac{1016q_{2}^{3}G\pi^{5}}{243\Omega_{(3)}\Omega_{(4)}q_{5}^{3}\sigma_{0}^{7/3}} T^{5} + \frac{64q_{2}G\pi^{3}}{27\Omega_{(3)}\Omega_{(4)}q_{5}\sigma_{0}^{7/3}}
T^{3}\Big)^{-1} \Big)}\Bigg)^{2}.
\end{eqnarray}
It may be noted that the scale factor are functions of the
tachyonic potential. Thus, as the
  tachyonic potential increases, the open
 universe expands to  infinity at $TA=\infty$, and the
 closed universe oscillates.

We can now write an expression for the energy density of the open and closed universes as
 \begin{eqnarray}
  \rho_{o/c,ac} &=& \frac{3}{8\pi l_{P}^{2}}(H_{o/c,ac}^{2}+k/a^{2})  \nonumber \\&\simeq&
   \frac{3}{8\pi l_{P}^{2}}H_{flat,ac}^{2} \nonumber \\ &\simeq&
    \frac{3}{8\pi L_{P}^{2}}\left(\frac{1}{V(TA)}\right)\left(\frac{1016q_{2}^{3}G\pi^{5}}{243\Omega_{(3)}\Omega_{(4)}q_{5}^{3}\sigma_{0}^{7/3}} T^{5} + \frac{64q_{2}G\pi^{3}}{27\Omega_{(3)}\Omega_{(4)}q_{5}\sigma_{0}^{7/3}}
T^{3}\right)^{2}\nonumber \\ &&\times  \left(\frac{8}{3}\alpha_{0}\sqrt{\frac{\pi\mu}{L_{P}^{2}}}\pm \sqrt{ \left(\frac{16}{3}\alpha_{0}^{2}\frac{\pi\mu}{L_{P}^{2}}+\frac{16\pi }{L_{P}^{2}}\left(\frac{1016q_{2}^{3}G\pi^{5}}{243\Omega_{(3)}\Omega_{(4)}q_{5}^{3}\sigma_{0}^{7/3}} T^{5} + \frac{64q_{2}G\pi^{3}}{27\Omega_{(3)}\Omega_{(4)}q_{5}\sigma_{0}^{7/3}}
T^{3}  \right)^{-1}   \right)} \right)^{2} \nonumber \\ &=&
\rho_{flat,ac}.
\label{mmmmm19}
\end{eqnarray}
It may be noted that this energy density also depends only on the
M2-M5 brane system, and not the type of the universe. Thus, the
 energy density for  all the three types of the universe is the same.

\section{Conclusion} \label{sum}
In this paper, we used the holographic cosmology conjecture to analyze the dynamics of a 6-D inflating spacetime..
The inflation started with the formation of a system of M2-M5 branes. The M5-brane was connected
to the anti-M5-brane by a M2-brane. This made it possible for the degrees of freedom to flow from
the anti-M5-brane to the M5-brane causing inflation. The inflation was driven by the difference in
the degrees of freedom between a region and the cosmological horizon surrounding that region. The inflation
ended with the separation of the M5-brane from the anti-M5-brane.
However, as the M5-brane approaches a second anti-M5-brane, at a later stage, tachyonic states where formed. These
tachyonic states opened a new bridge between the M5-brane and the anti-M5-brane. This again caused the degrees of
the freedom to flow into the M5-brane, and hence the expansion of the universe.
It may be noted that thermodynamic approach to gravity has also been studied using the generalized
uncertainty principle \cite{th}. In fact, holographic cosmology  with the
generalized uncertainty principle has
also been studied \cite{ht}. It would thus be interesting to generalize the results of this paper using the
generalized uncertainty principle.

\section*{Acknowledgments}
\noindent
Authors wish to thank  Douglas J Smith for useful discussions.
 The research of AFA is supported by Benha University (www.bu.edu.eg)‎. Also, the research of A.Sepehri is supported by
 Research Institute for Astronomy and Astrophysics of Maragha
(RIAAM), Maragha, Iran..

 \end{document}